\documentclass[%
twocolumn,10pt,
superscriptaddress,
 amsmath,amssymb,
 aps,pra
]{revtex4-1}

\pdfoutput=1

\usepackage{graphicx}
\usepackage{dcolumn}
\usepackage{bm}

\usepackage{braket}
\usepackage{xcolor}
\usepackage{units,ulem}
\usepackage{bbold}
\usepackage{multirow}

\begin{document}

\title{Experimental realization of equiangular three-state quantum key distribution}

\author{Matteo Schiavon}
\email{matteoschiav@yahoo.it}
\affiliation{Dipartimento di Ingegneria dell'Informazione,
Universit\`a di Padova, via Venezia 15, 35131 Padova}
\author{Giuseppe Vallone}%
\email{vallone@dei.unipd.it}
\affiliation{Dipartimento di Ingegneria dell'Informazione,
Universit\`a di Padova, via Venezia 15, 35131 Padova}
\affiliation{Istituto di Fotonica e Nanotecnologie, CNR, Padova, Italy}
\author{Paolo Villoresi}
\email{paolo.villoresi@dei.unipd.it}
\affiliation{Dipartimento di Ingegneria dell'Informazione,
Universit\`a di Padova, via Venezia 15, 35131 Padova}
\affiliation{Istituto di Fotonica e Nanotecnologie, CNR, Padova, Italy}

\date{\today}

\begin{abstract}
Quantum key distribution using three states in equiangular configuration combines a security threshold comparable with the one of the Bennett-Brassard 1984 protocol and a quantum bit error rate (QBER) estimation that does not need to reveal part of the key.
We implement an entanglement-based version of the Renes 2004 protocol, using only passive optic elements in a linear scheme for the measurement POVM, generating an asymptotic secure key rate of more than $\unit[10]{kbit/s}$, with a mean QBER of $1.6\%$.
We then demonstrate its security in the case of finite key and evaluate the key rate for both collective and general attacks.
\end{abstract}

\pacs{Valid PACS appear here}

\maketitle

\section{Introduction}
Quantum Key Distribution (QKD) allows two remote parties, Alice and Bob, to generate a secret shared string of bits, that can be used for symmetric cryptography or other cryptographic protocols.
Unlike classical key distribution, whose security is based on the computational difficulty of solving certain classes of problems, the security of quantum key distribution comes from the impossibility for an eavesdropper (Eve) to acquire information about an exchanged state without perturbing it.
This goal is obtained by using non-orthogonal states, as in the first QKD protocol, introduced by Bennett and Brassard in 1984 (BB84), that uses four different states of two mutually unbiased bases \cite{Bennett1984}.
In 1992, Bennett showed that two non-orthogonal states are sufficient for QKD \cite{Bennett1992}.
This protocol, however, has the drawback that the secure key rate is strongly affected by losses: indeed, Eve can extract information by increasing the losses and performing the so called unambiguous state discrimination (USD) attack~ \cite{Tamaki2003,Tamaki2004}.
The addition of a third state is sufficient to make the B92 protocol unconditionally secure independently from the noise in the quantum channel \cite{Fung2006,luca09pra}: however, rates comparable to the BB84 protocols can be obtained only when four states are detected at the receiver.

The optimal three-state QKD protocol, introduced in 2000 by Phoenix-Barnett-Chefles (PBC00)~\cite{Phoenix2000}, uses states that form an equilateral triangle in the X-Z plane of the Bloch sphere.
The symmetry of this protocol can indeed be exploited to obtain rates comparable with the BB84 protocol but requiring only three detectors instead of four.
This protocol, however, still requires the public exchange of part of the sifted key in order to estimate the QBER.
An improvement of this protocol, introduced by Renes in 2004 (R04)~\cite{Renes2004}, estimates the error rate from the number of inconclusive events, thus allowing to use all conclusive ones for key extraction.
The unconditional security of the PBC00 has been demonstrated, in the asymptotic case, in 2005 \cite{Boileau2005}, for a bit error rate of up to $\unit[9.81]{\%}$.
The security of the PBC00 has been demonstrated also in the case of finite key \cite{Mafu2014}, using the framework introduced by Scarani and Renner \cite{Scarani2008} and the postselection technique \cite{Christandl2009}.
Their work, however, does not consider the security of the R04 protocol, therefore excluding one of the most interesting features of this class of symmetrical codes.
Despite all these theoretical results, no experimental implementation of equiangular three state QKD protocols has been reported so far.
The measurement apparatus proposed in the original work implements the trine measurement using an interferometric setup at the receiver~\cite{Phoenix2000}.
This scheme requires careful alignment \cite{Clarke2001,Clarke2001-1} and is not assured to have the long term stability required by Quantum Key Distribution (a stability of about half an hour is reported in \cite{Clarke2001-1}).
A new experimental scheme implementing the trine measurement using only passive optical elements has been proposed in \cite{Saunders2012}.
Using this apparatus, we demonstrate the feasibility of equiangular three state QKD and assess its key generation rate, both in the asymptotic limit of infinite key and taking into account finite key effects.

\section{Results}
\subsection{Protocol}
\begin{figure}[t]
   \centering
      \includegraphics[width=0.7\linewidth]{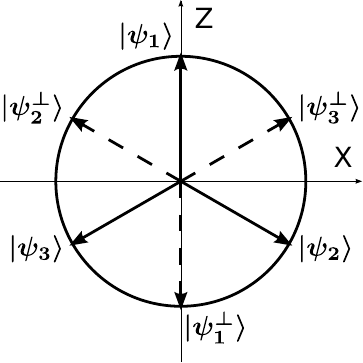}
      \includegraphics[width=0.7\linewidth]{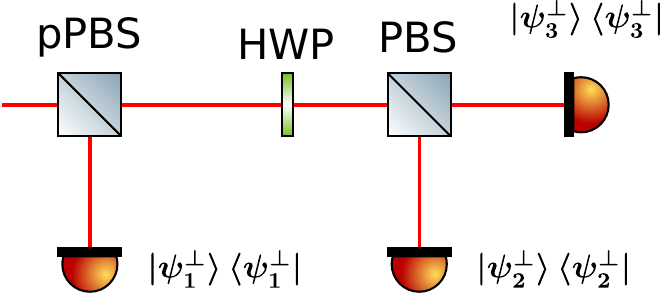}
      \caption{States used in the R04 protocol (above) and POVM $\{ \Pi_i \}$ used for the measurement (below). The states lie in the X-Z plane of the Bloch sphere. They are grouped into the sets $S_1 = \{ \ket{\psi_1}, \ket{\psi_2} \}$, $S_2 = \{ \ket{\psi_2}, \ket{\psi_3} \}$, and $S_3 = \{ \ket{\psi_3}, \ket{\psi_1} \}$, where the first element of each set corresponds to bit 0 and the second to bit 1. The POVM is implemented by using a partially polarizing beam-splitter (pPBS), a half-wave plate (HWP) at $22.5^\circ$ and a polarizing beam-splitter (PBS).}
      \label{fig:bloch}
\end{figure}

The R04 protocol uses three quantum states, $\{ \ket{\psi_1}, \ket{\psi_2}, \ket{\psi_3} \}$, placed in an equilateral triangle in the X-Z plane of the Bloch sphere, as shown in Figure \ref{fig:bloch}.
The states are grouped into three different sets, $S_1 = \{ \ket{\psi_1}, \ket{\psi_2} \}$, $S_2 = \{ \ket{\psi_2}, \ket{\psi_3} \}$, and $S_3 = \{ \ket{\psi_3}, \ket{\psi_1} \}$. 
In each set, the first state is associated with the bit $0$ and the second with the bit $1$.
Differently from other QKD protocols, each state brings no information about its associated bit before the information about the used set is disclosed. \\

We implemented an entanglement-based version of the protocol, using polarization-entangled photon pairs in the singlet state
\begin{equation}
				\ket{\Psi^{-}} = \frac{\ket{H}_A \ket{V}_B - \ket{V}_A \ket{H}_B}{\sqrt{2}},
				\label{eq:psi-}
\end{equation}
where the subscripts $A$ and $B$ indicate the photons going to, respectively, Alice's and Bob's detection apparatus.
In this state, photons A and B are anti-correlated in any measurement basis.
Alice measures photon $A$ using the POVM $\{\Pi_i\equiv\frac23\ket{\psi_i^\perp}\bra{\psi_i^\perp}\}$, with $\ket{\psi_1^\perp} = \ket{V}$, $\ket{\psi_2^\perp} = \frac{\sqrt{3}}{2}\ket{H} - \frac{1}{2}\ket{V}$, $\ket{\psi_3^\perp} = \frac{\sqrt{3}}{2}\ket{H} + \frac{1}{2}\ket{V}$. In our notation the state $\ket{\psi_i^\perp}$ is orthogonal to $\ket{\psi_i}$.
When Alice obtains a detection in the state $\ket{\psi_i^\perp}$ (each with probability $\frac{1}{3}$), she is sending to Bob the state $\ket{\psi_i}$ where  $\ket{\psi_1} = \ket{H}$, $\ket{\psi_2} = \frac{1}{2}\ket{H} + \frac{\sqrt{3}}{2}\ket{V}$, and $\ket{\psi_3} = \frac{1}{2}\ket{H} - \frac{\sqrt{3}}{2}\ket{V}$.
This operation corresponds to the random preparation, with equal probability, of one of the three states $\{\ket{\psi_1}, \ket{\psi_2}, \ket{\psi_3}\}$, as in the prepare-and-measurement scheme of the R04 described in \cite{Boileau2005,Scarani2009}.
Bob performs his measurements in the same POVM as Alice $\{ \Pi_i \}$.
After all measurements, Bob and Alice compare the
instants of their events, 
keeping only those where both have a detection within a fixed coincidence window. 

Even if they have already exchanged all symbols, Alice and Bob do not share any bit string yet, because each state can mean both 0 or 1.
Alice uses a QRNG to choose the bit value for each symbol.
The combination of the state and the bit value unambiguously determines the set $S_i$ used for that event (for example, if Alice sends $\ket{\psi_2}$ and the QRNG gives $1$, the set used for that event is $S_1$).
For each event, Alice tells Bob the corresponding set by sending him the value of the index $i$.
Bob uses $i$ to associate $\ket{\psi_2^\perp}$ (for $i = 1$), $\ket{\psi_3^\perp}$ (for $i = 2$), and $\ket{\psi_1^\perp}$ (for $i = 3$) with bit 0, and $\ket{\psi_1^\perp}$ (for $i = 1$), $\ket{\psi_2^\perp}$ (for $i = 2$), and $\ket{\psi_3^\perp}$ (for $i = 3$) with bit 1.
All other combinations are marked as inconclusive, since Bob is not able to determine the state sent by Alice.
Bob tells Alice which events are inconclusive and they both discard them.
They then estimate the quantum bit error rate (QBER) from the fraction of inconclusive events \cite{Renes2004,Boileau2005}, and use this information to distill the key using error correction and privacy amplification \cite{Scarani2008-1}. 

\subsection{Setup}
\begin{figure}[t]
   \centering
      \includegraphics[width=\linewidth]{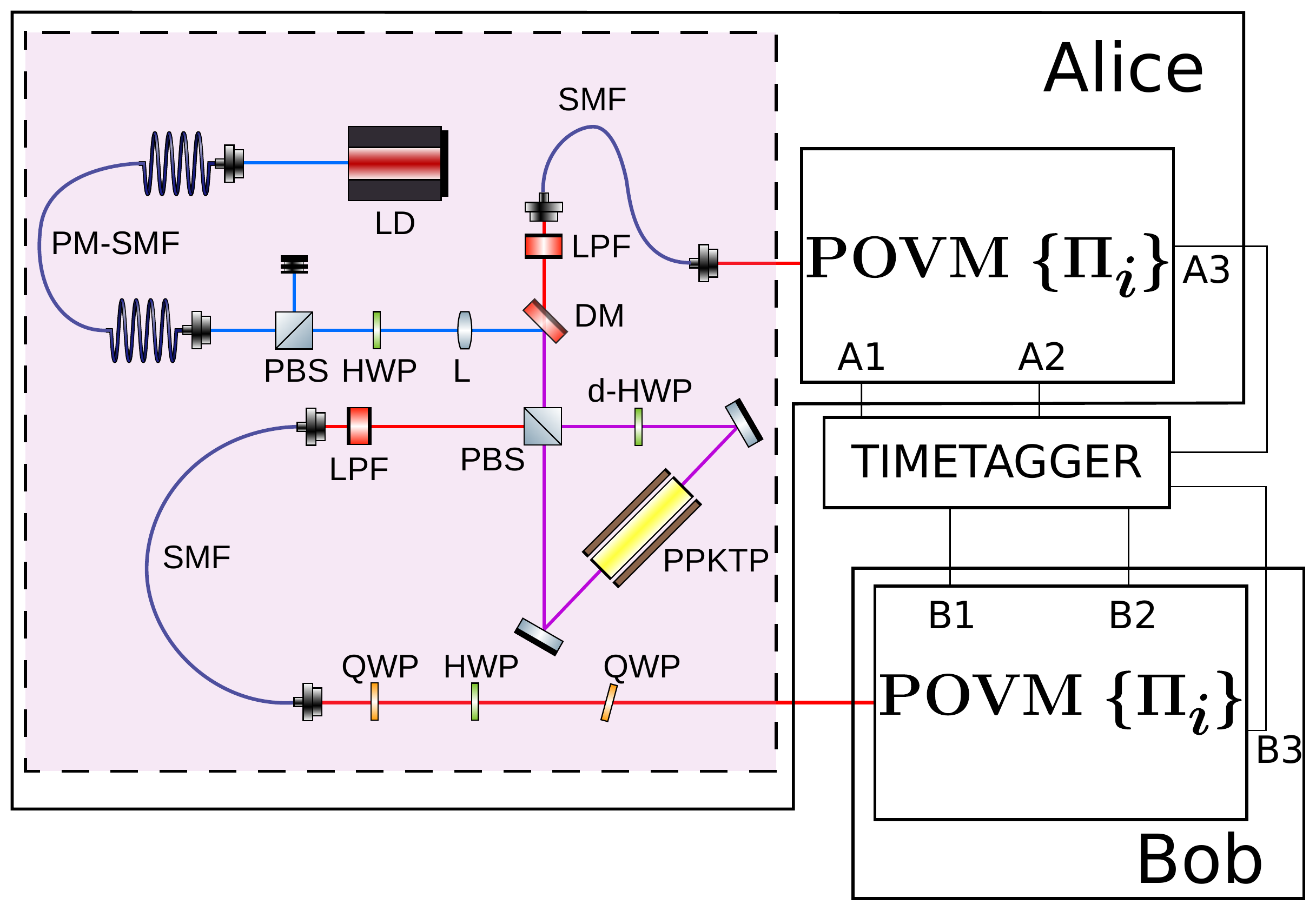}
      \caption{Experimental setup used for the experiment. The light issued by a laser diode (LD) at $\unit[404.5]{nm}$ is collected into a polarization maintaining single mode fiber (PM-SMF) for spatial mode filtering. The polarization is adjusted by a PBS followed by a HWP and the beam is focused by a $\unit[200]{mm}$ lens (L) into the center of the interferometer. The interferometer is formed by a PBS, a dual-wavelength HWP (d-HWP) and a PPKTP crystal, where the down-conversion takes place. Photons exiting the interferometer are collected into single mode fibers (SMF), preceded by a long-pass filter (LPF) to filter out residual pump intensity. Fiber birefringence is compensated by two quarter-wave plates (QWP) and a half-wave plate (HWP). The receiver consists in an implementation of the POVM $\{ \Pi_i \}$, with output Ai and Bi corresponding to $\ket{\psi_i^\perp}\bra{\psi_i^\perp}$ for, respectively, Alice and Bob. All the output signals are timed by a common timetagger.} 
\label{fig:setup}
\end{figure}

The experimental setup is shown in Figure \ref{fig:setup}.
Entangled photon pairs are produced by using a $\unit[30]{mm}$ periodically poled KTP crystal in a polarization-based Sagnac interferometer \cite{Kim2006,Fedrizzi2007}.
The source is pumped with a continuous wave (CW) laser at $\unit[404.5]{nm}$, with a power of $\unit[3.5]{mW}$.
The down-converted photons have a central wavelength of $\unit[809]{nm}$, with $\unit[0.2]{nm}$ full width at half maximum (FWHM), and are collected into single-mode fibers.
In this configuration, the setup has a mean coincidence rate of $\unit[29]{kHz}$, with a 5\% heralding ratio.
The fraction of multi-pair over one pair events, measured by putting one output of the source into a Hanbury-Brown-Twiss interferometer, is $\sim 3 \cdot 10^{-3}$.
Among these events, only those which are partially correlated in the polarization degree of freedom are exploitable by Eve through the photon number splitting (PNS) attack \cite{Scarani2008}.
The ratio of the number of these events to all multi-pair ones is $\zeta \simeq \frac{\tau_c}{\Delta t} = 5 \cdot 10^{-3}$, where $\tau_c = \unit[8]{ps}$ is the coherence time of down-converted photons and $\Delta t = \unit[1.5]{ns}$ is the coincidence window.
The fraction of correlated multi-photon events over the total number of detection events is $\sim 1.5 \cdot 10^{-5}$, thus the information leaked to Eve is negligible.
A set of two quarter-wave plates (QWP) and one half-wave plate (HWP) is placed at the exit of the fiber at Bob's side in order to compensate polarization rotations induced by fiber birefringence. 

The receiving apparatus implementing the POVM $\{\Pi_i\}$ consists of a partially polarizing beam-splitter (pPBS), that completely transmits the horizontal polarization and has a reflectivity of $\unit[66.7]{\%}$ for the vertical polarization, followed by a HWP at $\theta = 22.5^\circ$ and a polarizing beam-splitter (PBS).
Given an arbitrary input state $\ket{\phi} = \alpha \ket{H} + \beta \ket{V}$, with $|\alpha|^2 + |\beta|^2 = 1$, the pPBS routes it to detector $1$ with probability $P_1 = \frac{2}{3}|\beta|^2$.
The state at the transmitting output of the pPBS, $\alpha \ket{H} + \frac{1}{\sqrt{3}} \beta \ket{V}$, is transformed into $\alpha \ket{-} + \frac{1}{\sqrt{3}} \beta \ket{+} = \frac{1}{\sqrt{2}} ( \alpha + \frac{1}{\sqrt{3}} \beta) \ket{H} - \frac{1}{\sqrt{2}} ( \alpha - \frac{1}{\sqrt{3}} \beta) \ket{V}$ by the
HWP.
Thus, the probabilities of a click at detectors $2$ and $3$ are, respectively, $P_2 = \frac{1}{2} |\alpha - \frac{1}{\sqrt{3}} \beta |^2$  and $P_3 = \frac{1}{2} | \alpha + \frac{1}{\sqrt{3}} |^2$.
It is easy to check that the above
probabilities can be also written as $P_i = \bra{\phi} \Pi_i \ket{\phi}$, with $\Pi_i = \frac{2}{3} \ket{\psi_i^\perp} \bra{\psi_i^\perp}$: then, the above described apparatus implements the POVM $\{\Pi_i\}$.
Photons are detected using silicon single photon counting modules (SPCM), characterized by a dead time of $\unit[21]{ns}$ and a jitter of $\sim800$ ps FWHM. Detection events are time-tagged with a resolution of $\unit[81]{ps}$. 

\subsection{Data acquisition}
\begin{figure}[t]
   \centering
   \includegraphics[width=\linewidth]{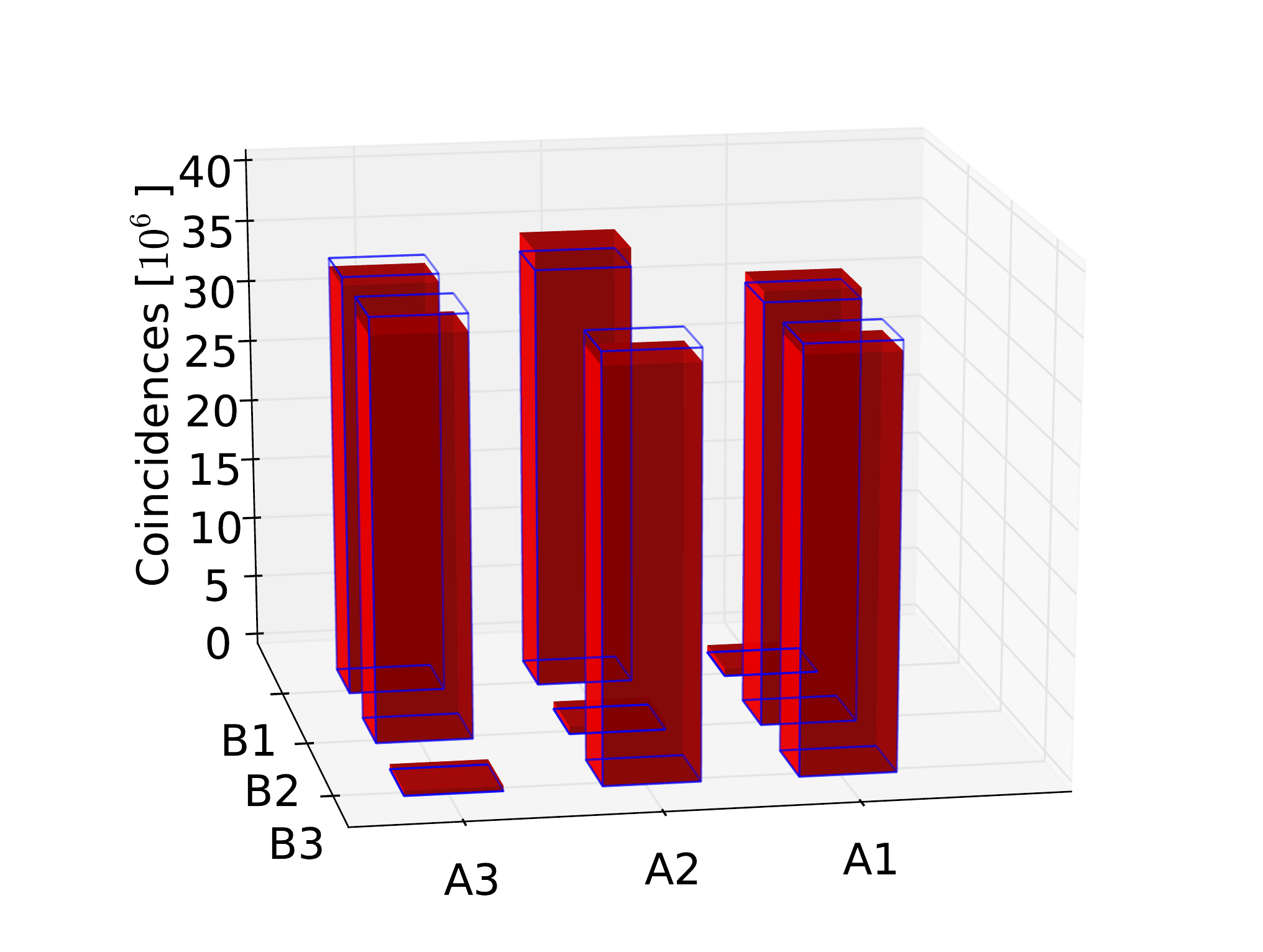}
   \caption{Total number of coincidences at the different detectors. Full (red) bars correspond to detected events and (blue) contours represent the expected number of detection events. }
   \label{fig:histogram}
\end{figure}

A two hour continuous run of the apparatus has led to the exchange of about $10^{9}$ symbols within a coincidence window of $\unit[1.5]{ns}$.
The events can be described as pairs (Ai,Bj), with $i$ the number of the detector clicking at Alice's side and $j$ the one at Bob's side.
Event distribution is shown in table \ref{tab:data} and Figure \ref{fig:histogram}.
\begin{table}[h]
   \centering
   \begin{tabular}{|c|c|c|c|}
      \hline
      & A1 & A2 & A3 \\
      \hline
      B1 & 0.6 & 35.8 & 33.6 \\
      B2 & 35.1 & 0.6 & 32.8 \\
      B3 & 33.4 & 33.2 & 0.4 \\
      \hline
   \end{tabular}   
   \caption{\label{tab:data} Total number of coincidences at the different detectors (million events). The cell (Ai,Bj) corresponds to a coincidence of Alice's detector $i$ and Bob's detector $j$.}
\end{table}
After the collection of all data, a QRNG \cite{Vallone2014} is used to generate the bit value for each symbol.
Coincidence events are then analyzed using the sifting procedure summarized in Table \ref{tab:sifting}.
\begin{table}[h]
   \centering
   \begin{tabular}{|c|c|c|c||c|c|c|}
      \hline
      & \multicolumn{3}{c||}{Bit = 0} & \multicolumn{3}{c|}{Bit = 1} \\
      \hline
      & A1 & A2 & A3 & A1 & A2 & A3\\
      \hline
      B1 & 1 & Inc & 0 & 0 & 1 & Inc \\
      B2 & 0 & 1 & Inc & Inc & 0 & 1 \\
      B3 & Inc & 0 & 1 & 1 & Inc & 0\\
      \hline
   \end{tabular}
   \caption{\label{tab:sifting} Sifting procedure, according to the random choice of the bit at Alice's side (on the left for 0, on the right for 1). The cell (Ai,Bj) stands for a coincidence between Alice's detector $i$ and Bob's detector $j$. Inconclusive events are marked as ``Inc''. The events in the diagonal (Ai,Bi) give an error independently from the bit choice. The other combinations (Ai,Bj), with $i \neq j$, are either a ``good'' conclusive or an inconclusive event, according to Alice's choice.}
\end{table}
The events of the form (Ai,Bi) are bit errors, while the others are either a ``good'' conclusive or an inconclusive result according to Alice's choice.
The string of conclusive results gives the sifted key, from which a secret key can be distilled using classical post-processing. 

\subsection{Secret key rate}
Post-processing consists of a series of passages that transform a partially correlated, partially secret key into a new one Eve has negligible information of \cite{Scarani2008-1,Scarani2009}.
The effect of these tasks is a reduction of the number of bits and can be quantified using the secret fraction $r$, defined as the ratio between secure and conclusive bits \cite{Scarani2009}.
In the asymptotic limit of infinitely long key, the key fraction of the R04 is given by \cite{Boileau2005}
\begin{equation}
   r = 1 - f_{EC} h(Q) - h\left({\frac{5}{4}Q}\right),
   \label{eq:asymp_r}
\end{equation}
where $f_{EC} = 1.1$ is the efficiency of the error correction protocol \cite{Mateo2015}.
The number of secure bits is given by $N_{conc} r$ and, dividing it by the exposure time, the secret key rate is obtained. 

\begin{figure}[t]
   \centering
   \includegraphics[width=\linewidth]{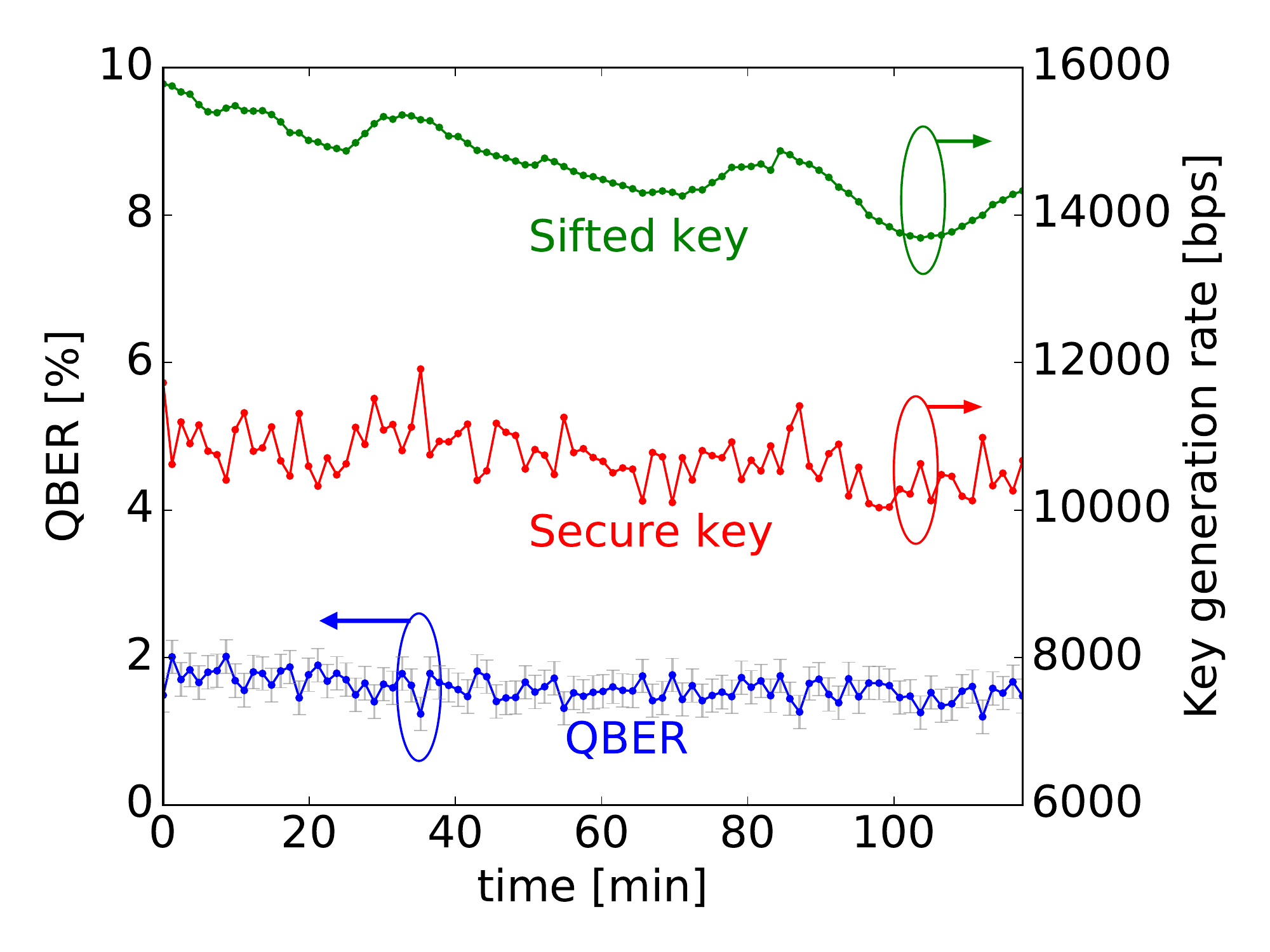}
	 \caption{Result obtained during 2 hour of continue acquisition. The time has been divided into $90$ blocks of about $\unit[80]{s}$ each, with a mean number of $1.1 \cdot 10^{6}$ sifted bits. The QBER is estimated from inconclusive events, with Poissonian error bars. The secret key rate is estimated for the case of infinitely long key, using equation (\ref{eq:asymp_r}).}
   \label{fig:keyrate}
\end{figure}

Figure \ref{fig:keyrate} shows the behavior of the QBER and of the key rate during a two hour acquisition.
The slight reduction in the sifted key rate is probably due to a misalignment in the fiber coupling of the entangled source.
The losses can be estimated from the ratio of coincidences over single counts.
The measured $5\%$ heralding efficiency corresponds to a total loss level of $\unit[13]{dB}$, with a contribution of $\unit[1.5]{dB}$ due to the POVM.
The QBER is estimated as $Q = \frac{1-2I}{1-I}$, where $I$ is the fraction of inconclusive results \cite{Boileau2005}.
The QBER remains almost constant at a level below $2\%$ for all the acquisition, thus confirming the stability of both the source and the POVM $\{ \Pi_i \}$ during the acquisition.
The visibility of the source in two mutually unbiased bases at the exit of the Sagnac interferometer, before fiber injection, has been measured to be between $97\%$ and $98\%$: then the measured QBER level can be attributed almost completely to the source.
A small contribution to the QBER is due to the small imbalances between the different channels of the POVM $\{ \Pi_i \}$.
These values, estimated by computing the ratio between the raw counts at the different detectors, vary between $0.95$ and $1.05$, in line with what observed in the previous implementation of the POVM \cite{Saunders2012}.

In a real scenario, the number of exchanged signals is always finite and the security analysis must take this fact into account.
The finite key analysis of the R04 protocol is very similar to the one of the PBC00 \cite{Mafu2014}, the only substantial difference lying in the estimation of the bit error rate.
\begin{figure}[t]
\includegraphics[width=\linewidth]{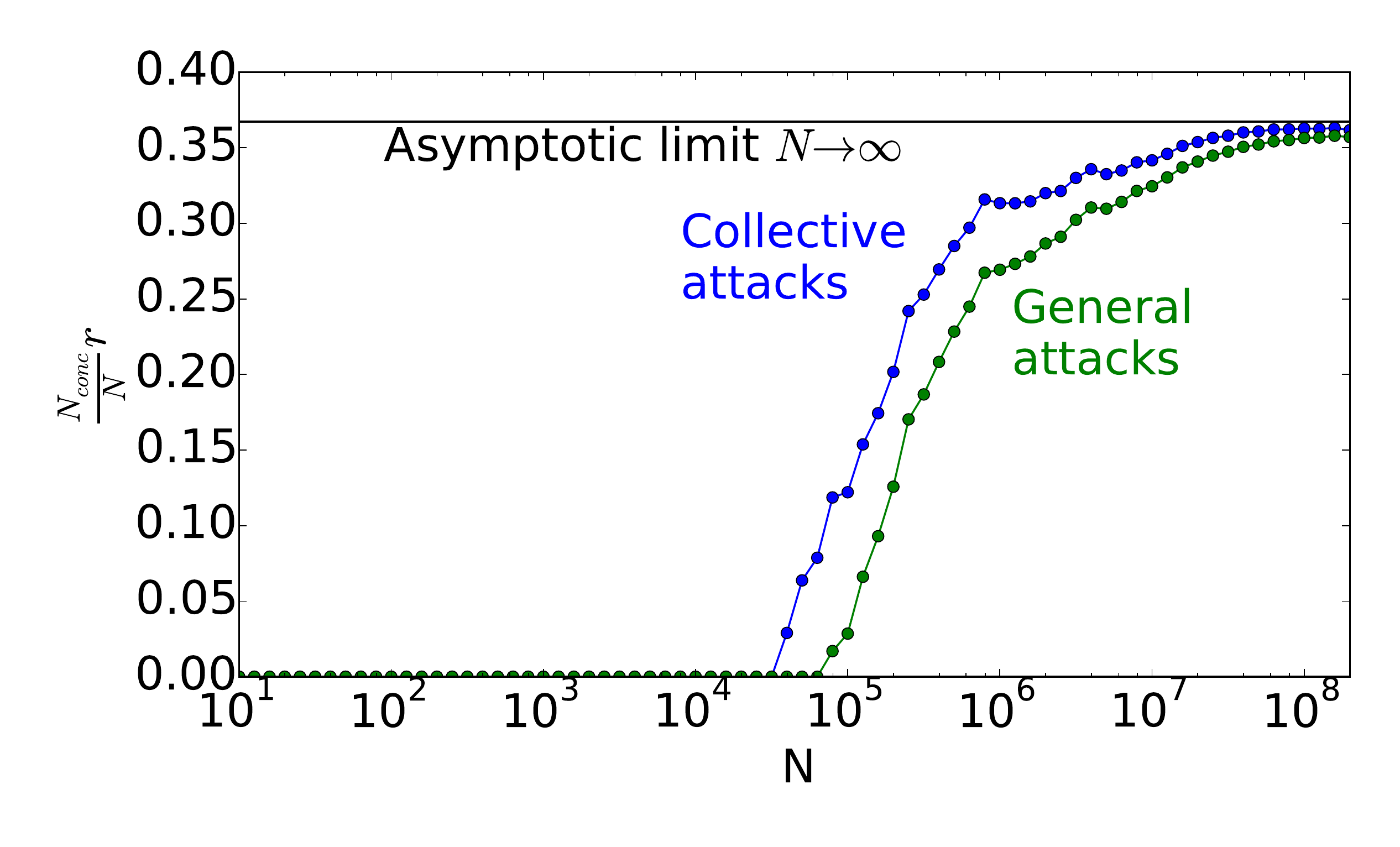}
\caption{Finite key analysis of the R04 protocol.
The y-axis represents the fraction of exchanged symbols giving a secure bit. Each point is calculated using the first N collected data for the estimation of the QBER. The security parameter for both collective and general attacks is fixed to $4 \cdot 10^{-10}$.}
\label{fig:finite}
\end{figure}
The PBC00 estimates it by comparing part of the sifted key through the public channel, while the R04 use the fraction of inconclusive results.
The method is based on the fact that the choice between ``good'' conclusive and inconclusive results is given by a random event at Alice's side after the exchange of all the qubits, therefore Eve has no way of differentiate between the two and their number is approximately equal \cite{Boileau2005}.
Defining $I$ the fraction of inconclusive results, the fraction of ``good'' conclusive ones can be written as $(1-Q)(1-I)$.
Using the Hoeffding bound \cite{Hoeffding1963}, the inequality
\begin{equation}
				\left| (1-Q)(1-I) - I \right| \leq \xi(\epsilon,N) := \sqrt{\frac{2}{N}\log\frac{2}{\epsilon}}
				\label{eq:hoeffding}
\end{equation}
is valid with probability at least $1-\epsilon$.
This implies that, with the same probability, the QBER $Q$ is less than $\widetilde{Q} := \frac{1-2I+\xi(\epsilon,N)}{1-I}$.
The secret key fraction, in the case of collective attacks, is \cite{Scarani2008,Mafu2014}
\begin{align}
				r_{col} = & \left[ 1 - h\left( \frac{5}{4} \widetilde{Q} \right) \right] - 7 \sqrt{\frac{1}{N} \log_2{\frac{2}{\bar{\epsilon}}}} \nonumber \\
				    & - \frac{1}{N}\log_2{\frac{1}{\epsilon_{EC}}} - \frac{1}{N}\log_2{\frac{2}{\epsilon_{PA}}} - f_{EC} h(Q)
				\label{eq:r_finite}
\end{align}
for $N$ exchanged signals, with security parameter $\epsilon_{col} := \bar{\epsilon} + \epsilon_{EC} + \epsilon_{PA} + \epsilon_{PE}$.
This result can be extended to the case of general attacks by exploiting the postselection technique \cite{Christandl2009}, giving \cite{Mafu2014} 
\begin{equation}
r_{gen} = r_{col} - \frac{6 \log_2{(N+1)}}{N},
\label{eq:r_finite_gen}
\end{equation}
with security parameter $\epsilon_{gen} = (N+1)^{3} \epsilon_{col}$. 

Figure \ref{fig:finite} shows the secret key fraction of the R04 protocol in the finite key scenario.
For each point, both the QBER and the number of conclusive events is evaluated on the first $N$ exchanged symbols.
The secret key fraction is calculated by using equations (\ref{eq:r_finite}) and (\ref{eq:r_finite_gen}).
For collective attacks, the chosen security parameter is $\epsilon_{col} = 4 \cdot 10^{-10}$, with $\bar{\epsilon} = \epsilon_{EC} = \epsilon_{PA} = \epsilon_{PE} = 10^{-10}$.
The same value has been chosen for $\epsilon_{gen}$, therefore the term $r_{col}$ of equation (\ref{eq:r_finite_gen}) is calculated using $\bar{\epsilon} = \epsilon_{EC} = \epsilon_{PA} = \epsilon_{PE} = \frac{10^{-10}}{(N+1)^3}$.
The plots show that at least $10^{4}$ - $10^{5}$ signals are necessary to exchange a key, while already $N = 10^{6}$ (slightly more than half a minute at $\unit[29]{kHz}$) gives a reasonable key fraction.
The difference between the key fraction for collective and general attacks is more marked for lower values of $N$ and tends to disappear for large number of exchanged symbols, where both approach the asymptotic key fraction. 

\section{Discussion}
In this work, we demonstrated the experimental feasibility of the equiangular three state QKD protocol R04.
We showed that the scheme proposed by \cite{Saunders2012} for the POVM is suitable for applications in Quantum Key Distribution.
State preparation was simplified by using an entanglement-based version of the protocol, with the same POVM at both Alice's and Bob's side.
We also showed that the estimation of the bit error rate from inconclusive results is feasible in the finite key scenario.
The implemented scheme was demonstrated to be stable and highly reliable, allowing a two hour data acquisition without any significative change in the QBER value.
The performance of the protocol is comparable with the BB84, despite the less efficient parameter estimation, both in the asymptotic limit and for finite key.
Its simpler receiving apparatus, requiring only three single photon detectors, makes it a valid alternative to current implementations of QKD based on the BB84 protocol.
Finally, this work extends the experimental investigation of equiangular spherical codes to a still uncovered area of quantum information: quantum key distribution.

\acknowledgements
We thank Davide G. Marangon and Nicola Laurenti for helpful discussions.
This work has been carried out within the Strategic-Research-Project QUANTUMFUTURE of the University of Padova and the Strategic-Research-Project QUINTET of the Department of Information Engineering, University of Padova.
M.S. acknowledges the Ministry of Research and the Center of Studies and Activities for Space (CISAS) ``Giuseppe Colombo'' for financial support.



\begin{thebibliography}{1}
\providecommand{\url}[1]{\texttt{#1}}
\providecommand{\urlprefix}{URL }
\providecommand{\eprint}[2][]{\url{#2}}


\bibitem{Bennett1984}
C.~H.~Bennett and G.~Brassard in \textit{Proceedings IEEE Int. Conf. on Computers, Systems and Signal Processing, Bangalore, India, 1984} (IEEE, New York), p. 175.


\bibitem{Bennett1992}
C.~H.~Bennett, Phys. Rev. Lett. \textbf{68}, 3121 (1992).

\bibitem{Tamaki2003}
K.~Tamaki, M.~Koashi, and N.~Imoto, Phys. Rev. Lett. \textbf{90}, 167904 (2003); 

\bibitem{Tamaki2004}
K.~Tamaki and N,~L\"utkenhaus, Phys. Rev. A \textbf{69}, 032316 (2004).

\bibitem{Fung2006}
C.-H.~F.~Fung and H.-K.~Lo, Phys. Rev. A \textbf{74}, 042342 (2006).

\bibitem{luca09pra}
M.~Lucamarini, G.~Di~Giuseppe, K.~Tamaki, 
Phys. Rev. A {\bf 80}, 32327 (2009).

\bibitem{Phoenix2000}
S.~Phoenix, S.~Barnett, and A.~Chefles, J. Mod. Opt. \textbf{47}, 507 (2000).

\bibitem{Renes2004}
J.~M.~Renes, Phys. Rev. A \textbf{70}, 052314 (2004).

\bibitem{Boileau2005}
J.-C.~Boileau, K.~Tamaki, J.~Batuwantudawe, R.~Laflamme, and J.~M.~Renes, Phys. Rev. Lett. \textbf{94}, 040503 (2005).

\bibitem{Mafu2014}
M.~Mafu, K.~Garapo, and F.~Petruccione, Phys. Rev. A \textbf{90}, 032308 (2014).

\bibitem{Scarani2008}
V.~Scarani and R.~Renner, Phys. Rev. Lett. \textbf{100}, 200501 (2008).

\bibitem{Christandl2009}
M.~Christandl, R.~K\"onig, and R.~Renner, Phys. Rev. Lett. \textbf{102}, 020504 (2009).

\bibitem{Clarke2001}
   R.~B.~M.~Clarke, A.~Chefles, S.~M.~Barnett, and E.~Riis, Phys. Rev. A \textbf{63}, 040305(R) (2001).

\bibitem{Clarke2001-1}
   R.~B.~M.~Clarke, V.~M.~Kendon, A.~Chefles, S.~M.~Barnett, E.~Riis, and M.~Sasaki, Phys. Rev. A \textbf{64}, 012303 (2001).

\bibitem{Saunders2012}
   D.~J.~Saunders, M.~S.~Palsson, G.~J.~Pryde, A.~J.~Scott, S.~M.~Barnett, and H.~M.~Wiseman, New J. Phys. \textbf{14}, 113020 (2012).

\bibitem{Scarani2009}
V.~Scarani, H.~Bechmann-Pasquinacci, N.~J.~Cerf, M.~Du\v{s}ek, N.~L\"utkenhaus, and M.~Peev, Rev. Mod. Phys. \textbf{81}, 1301 (2009).

\bibitem{Scarani2008-1}
V.~Scarani and R.~Renner in \textit{Proceedings of TQC2008}, Lecture Notes in Computer Science 5106 (Springer Verlag, Berlin, 2008), pp. 83-95. 

\bibitem{Kim2006}
T.~Kim, M.~Fiorentino, and F.~N.~C.~Wong, Phys. Rev. A \textbf{73}, 012316 (2006).

\bibitem{Fedrizzi2007}
A.~Fedrizzi, T.~Herbst, A.~Poppe, T.~Jennewein, and A.~Zeilinger, Opt. Express \textbf{15}, 15377 (2007).

\bibitem{Vallone2014}
G.~Vallone, D.~G.~Marangon, M.~Tomasin, and P.~Villoresi, Phys. Rev. A \textbf{90}, 52327 (2014).

\bibitem{Mateo2015}
J.~Martinez-Mateo, C.~Pacher, M.~Peev, A.~Ciurana, and V.~Martin, Quantum Information \& Computation, vol. 15, no. 5\&6, pp. 453-477 (2015).

\bibitem{Hoeffding1963}
W.~Hoeffding, J. Amer. Stat. Assoc. \textbf{58}, 13 (1963).

\end{thebibliography}
\end{document}